\documentclass[twocolumn, showpacs, prl, superscriptaddress]{revtex4}
\usepackage{amsmath}
\usepackage{graphicx}
\usepackage{subfigure}
\usepackage{hyperref}

\def \be {\begin{equation}}
\def \ee {\end{equation}}

\begin{document}
\title{Effect of Dilution on Spinodals and Pseudospinodals}

\author{K. Liu}
\email{kangliu@physics.bu.edu}
\affiliation{Department of Physics, Boston University, Boston, Massachusetts 02215}
\author{W. Klein}
\email{klein@bu.edu}
\affiliation{Department of Physics, Boston University, Boston, Massachusetts 02215}
\affiliation{Center for Computational Science, Boston University, Boston, Massachusetts 02215}
\author{C. A. Serino}
\email{cserino@physics.bu.edu}
\affiliation{Department of Physics, Boston University, Boston, Massachusetts 02215}

\date{\today}
\begin{abstract}


We investigate the effect of quenched dilution on the critical and spinodal points in the infinite range (mean-field) and long-range (near-mean-field) Ising model. We find that unlike the short-range Ising model, the effect of the dilution is not simply related to the divergence of the specific heat, i.e., the Harris criterion.  We also find the mean-field behavior differs from that of $d=4$ (upper critical dimension) nearest neighbor model at the critical point. These results are an important first step for understanding the effect of the spinodal on the nucleation process as well as the properties of the metastable state in systems of considerable interest in material science, geophysics, and econophysics which in general have defects. 
\end{abstract}


\maketitle

Long range interactions are common as well as important in both physical and social systems.  
Metals~\cite{klein-lookman,Run-Klein}  and earthquake faults~\cite{serino}, both with long-range elastic interactions, as well as biological and polymer materials~\cite{binder} all have processes strongly influenced by the long range nature of their interactions. In addition, economic systems as well as societal interactions that facilitate the transmission of diseases~\cite{dodds} have effective long range ``interactions'' due to globalization. In all of these systems metastable states are common~\cite{ball, kelton} and nucleation plays an important role. In addition, it has been established that in long range systems the nucleation process is strongly affected by the pseudospinodal~\cite{klein2007}. It is important to distinguish mean-field (mf) systems with infinite range interactions~\cite{kac,leb} from systems with long but finite range interactions. We will refer to the latter  as near-mean-field (nmf). Mean-field systems have spinodals while nmf systems have pseudospinodals~\cite{klein2007, gulba, heer-kl-st}. Pseudospinodals behave  as spinodals unless they are approached too closely as we will discuss below.

Although the spinodal and pseudospinodal have a strong effect on the metastable state and nucleation in
defect free mf and nmf systems~\cite{binder, klein2007}, the systems mentioned above generally have defects. In materials such as metals, the defects can take the form of impurities and/or vacancies~\cite{kelton}. In economic and other social systems agents that behave differently than the majority can be thought of as defects. Examples  include the presence of immune individuals in models of disease transmission~\cite{ball} and agents that do not participate in wealth exchange in economic models. This raises an important, but as yet unanswered, question: How, if it all, does the presence of defects affect the spinodal and pseudospinodal? This question is  analogous to   what was asked about critical points in systems with defects and short range interactions several decades ago. In the 70's, Harris~\cite{harris74} developed a criterion associated with the specific heat exponent in the defect free, or pure, system that predicted whether the critical point remaines well defined in short range Ising models in the presence of vacancies. In this paper we will investigate the same question for mf and nmf critical points, spinodals and pseudospinodals. As we will show, mf and nmf Ising systems have a  more complicated response to defects than nearest neighbor models. Moreover, the response is different than what one finds in $d=d_c=4$, the Ising upper critical dimension~\cite{ballesteros,ma}.

We begin by noting that the Ginzburg criterion for mf to be a good approximation in spin systems is that the fluctuations in the magnetization are small compared to its mean value. At the critical point~\cite{klein2007}
\be 
\label{ginz-cp}
G^{-1} = {\xi^{d}\chi_{T}\over \xi^{2d}\phi^{2}} = 0 \rightarrow G = R^{d}\epsilon^{2-d/2}\rightarrow \infty,
\ee
where for interactions with range $R$, the correlation length is $\xi = R\epsilon^{-1/2}$, the reduced temperature is $\epsilon = (T-T_{c})/T_{c}$, and $T_{c}$ is the critical temperature. The susceptibility, $\chi_{T} \propto \epsilon^{-1}$, and the magnetization per spin, $\phi\propto \epsilon^{1/2}$. We can use $G$ to distinguish between mf ($G\rightarrow \infty$) and nmf ($G\gg1$ but not infinite) for all $d$.

Near spinodals or pseudospinodals the same considerations lead to the criterion~\cite{klein2007} 
\be
\label{ginz-spin}
G_{s} = R^{d}\Delta h^{3/2 - d/4}\gg1
\ee
for there to be a well pronounced pseudospinodal; $G_{s}\rightarrow\infty$ for a true spinodal~\cite{klein2007, gulba}. Here the correlation length is $\xi = R\Delta h^{-1/4}$, the susceptibility is $\chi_{T}\sim \Delta h^{-1/2}$, and the difference between the magnetization and its value at the spinodal is $\Delta \phi\sim \Delta h^{1/2}$ where $\Delta h = (h_{s}-h)/h_{s}$ where $h_{s}$ is the value of the magnetic field at the spinodal. 
Some of the simulations we discuss below involve fully connected models, i.e., every spin interacts with every other spin. As in finite size scaling, we take the volume of the system, $N$, to be either $\xi^{d} = R^{d}\epsilon^{-d/2}$ (critical point) or $R^{d}\Delta h^{-d/4}$ (spinodal). Then $G = N\epsilon^{2}$(critical point) and $G_{s}=N\Delta h^{3/2}$ (spinodal).

The Harris criterion~\cite{harris74} (HC) compares the width of the distribution of vacant sites ($p\equiv$ fraction of occupied sites)  
to the distance from the  continuous transition. At the critical point this leads to no change in exponents if
\be
\label{harris}
{[\xi^{d}p(1-p)]^{1/2}\over \xi^{d}} \ll \epsilon\,.
\ee
Using again the fact that $\xi=R\epsilon^{-1/2}$ Eq.~\eqref{harris} results in
$G = R^{d}\epsilon^{2 - d/2} \gg 1$.

\begin{figure}
\centering
\includegraphics[width = \columnwidth]{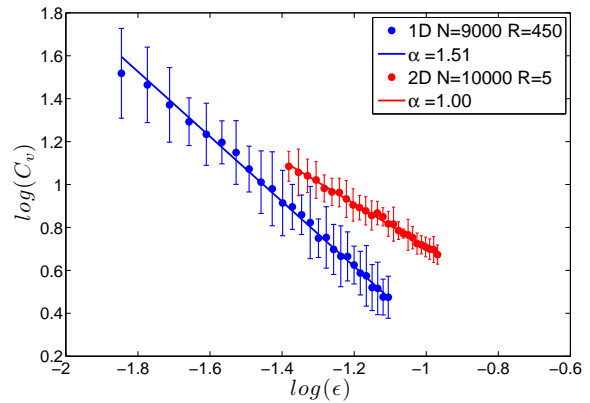}
\caption{(color online) Plots of the fluctuations in the energy(specific heat) as a function of the distance from the critical point for the pure system in $d=1$ and 2 for $T>T_{c}$.}
\label{fig:Fig1}
\end{figure}

How does the HC relate to the specific heat in mf and nmf systems?  
We can obtain the specific heat by calculating the energy fluctuations per unit volume. In mf and nmf, the energy fluctuations are the square of the magnetization fluctuations. 
The magnetization fluctuations per unit volume in mf and nmf do not scale as $\epsilon^{\beta}$ (with $\beta = 1/2$ at mean-field critical points) but rather as~\cite{klein2007} 
\be
\label{mag-scale}
m\sim {\epsilon^{1/2}\over (R^{d}\epsilon^{2-d/2})^{1/2}}\,.
\ee
At critical points, the scaling in Eq.~\eqref{mag-scale} can be seen by looking at a Landau-Ginzburg-Wilson Hamiltonian (LGW) in zero field~\cite{ma}
We scale all lengths with $\xi$ and $m({\vec x})$ by $\epsilon^{1/2}$ and obtain
\be
\label{lgw-scale}
H(m) = R^{d}\epsilon^{2-d/2}\!\!\int\! d{\vec y}\, \bigg [(\nabla \phi({\vec y}))^{2} + \phi^{2}({\vec y}) + \frac{\phi^4({\vec y})}{{\left(R^d\epsilon^{2-d/2}\right)}^2}\bigg ]\,,
\ee
where ${\vec y}$ is the scaled length and $\phi({\vec y})$ is the scaled magnetization. As can be readily seen from using  Eq.~\eqref{lgw-scale} to calculate the partition function, $\phi({\vec y})$ is of order  $(R^{d}\epsilon^{2-d/2})^{-1/2}$. This scaling reduces the LGW Hamiltonian to a Gaussian because we can ignore the scaled $\phi^{4}$ term.
For $T>T_{c}$ and $h=0$, the specific heat at constant $h$ is
\be
\label{sh>Tc}
C_{V} \sim \bigg ({\epsilon^{1/2}\over \sqrt{R^{d}\epsilon^{2-d/2}}}\bigg )^{\!4}R^{d}\epsilon^{-d/2} = {1\over R^{d}\epsilon^{2-d/2}}\,.
\ee
This is the square of the energy fluctuations per unit volume and hence for $T>T_{c}$, the HC for the mf (or nmf) critical point to be unaffected by the dilution becomes $C_{V}\ll1$. In Fig.~\ref{fig:Fig1} we plot the energy fluctuations (specific heat) of the non-diluted model as a function of $\epsilon$ in $d=1, 2$ (blue and red, respectively). For $T>T_{c}$ the divergence is consistent with Eq.~\eqref{sh>Tc}, namely $\epsilon^{-3/2}$ in $d=1$ and $\epsilon^{-1}$ in $d=2$.

There are several points to be made. The first is that at the critical point the nmf criterion that $G\gg1$ is the same as the HC so that if the system is well approximated by mf theory, the critical point remains well defined and unchanged in the presence of vacancies.  In addition, for $G\rightarrow\infty$ (mf) the critical point is unchanged by the dilution.
The second point is that for $R\gg1$ but finite, the affect of the dilution depends on the distance from the critical point $\epsilon$. Hence dilution in the $d=3$ Ising model  has no affect on the apparent mean-field critical point if $G\gg1$ and finite. However, as $\epsilon$ decreases and the system crosses over to $d=3$ short-range Ising behavior, the vacancies alter the critical behavior from what one obtains in the pure model.

Our result is somewhat different than one obtained by  Ballesteros {\it et al.}~\cite{ballesteros} who looked at the nearest neighbor dilute Ising model in $d=d_c=4$ where the specific heat exponent $\alpha = 0$, so the HC is indeterminate~\cite{harris74}. Despite no power-law divergence, the pure system does have a logarithmically diverging specific heat. It was found in Ref.~\cite{ballesteros} that the logarithmic corrections to the exponents changed; however, the exponents themselves did not. For the mf ($G\rightarrow \infty$) system we consider, the HC predicts that the nature of the critical point remains the same as the pure system. 

 \begin{figure}
\centering
\includegraphics[width = \columnwidth]{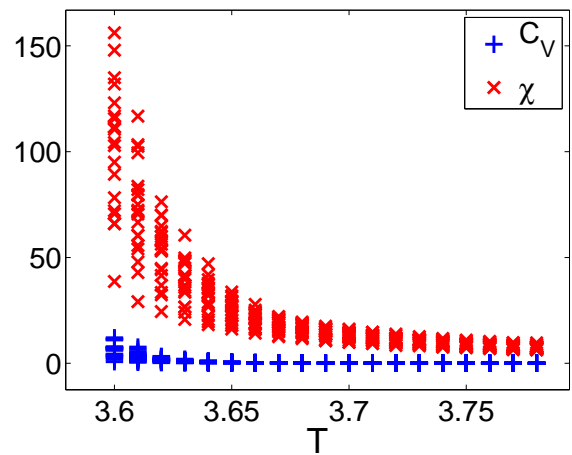}
\caption{(color online) Susceptibility and specific heat as a function of $\epsilon$ for a system with average dilution of 10\%. Sites were designated as empty with a 0.1 probability. Twenty systems were simulated.}
\label{fig:Fig2}
\end{figure}


In Fig.~\ref{fig:Fig2} we plot the susceptibility $\chi$ and specific heat $C_{V}$ as a function of $\epsilon>0$ for the system with dilution and $G\gg1$ but finite for the fully connected model. As we approach the critical point and $G$ decreases, the spread in both $C_{V}$ and $\chi$ increases consistent with the HC. (See Fig.~\ref{fig:Fig5})

We next turn  to the spinodal. The HC is
\be
\label{harris-sp}
{[\xi^{d}p(1-p)]^{1/2}\over \xi^{d}} = {\text{constant}\over \xi^{d/2}} \ll \Delta h\,.
\ee
With $\xi = R\Delta h^{-1/4}$, the HC becomes
\be
\label{harris-sp2}
{\Delta h^{-1/2}\over R^{d}\Delta h^{3/2-d/4}} \ll 1\,.
\ee
The specific heat is again related to the fluctuations of the energy. Near the spinodal the magnetization  is   non-zero. 
By using essentially   the same argument as outlined in Eq.~\eqref{lgw-scale}, the fluctuations in the magnetization scale as 
\be
\label{fluct-sp}
m \sim {\Delta h^{1/2}\over \sqrt{R^{d}\Delta h^{3/2-d/4}}},
\ee
where we have expanded the LGW Hamiltonian around the magnetization at the spinodal~\cite{klein2007}.

\begin{figure}
\begin{center}
\includegraphics[width = \columnwidth]{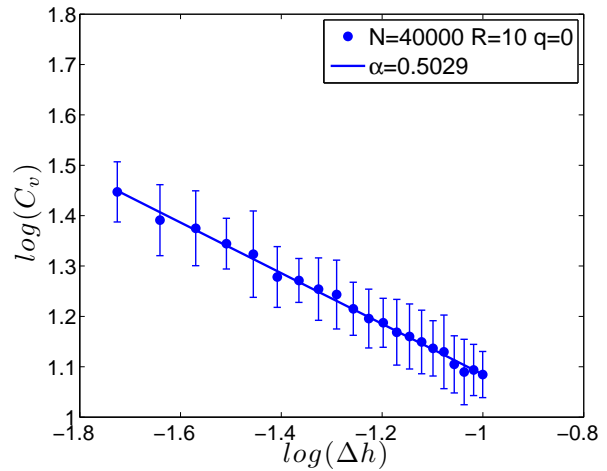}
\caption{The specific heat for the pure system  plotted as a function of $h$ for $d=2$. The fit is consistent with a divergence with exponent 1/2 consistent with the scaling analysis.} 
\label{fig:Fig3}
\end{center}
\end{figure}

The specific heat (energy fluctuations) at the spinodal is 
\begin{align}
C_{SP} &\propto \bigg [\bigg ({\Delta h^{1/2}\over \sqrt{R^{d}\Delta h^{3/2-d/4}}} + \phi\bigg )^{\!2} - \phi^{2}\bigg ]^{2}R^{d}\Delta h^{-d/4}\nonumber\\
&\propto \Delta h^{-1/2}\,.
\label{sh-sp}
\end{align}
Note that near the spinodal the magnetization does not go to zero as the spinodal is approached. Near the pseudospinodal, the HC is not simply related to the specific heat. What is required for the pseudospinodal to be unchanged by the presence of dilution according to the HC is that the specific heat divided by the Ginzburg parameter be small. This result implies that in mean-field where the $R\rightarrow \infty$ before $\Delta h\rightarrow 0$~\cite{kac,leb} the spinodal is unaffected by the dilution.
In Fig.~\ref{fig:Fig3}  we plot the specific heat (energy fluctuations) in the long-range Ising model without dilution as the spinodal is approached. The result is consistent with $C_{\rm SP}\sim \Delta h^{-1/2}$. 

 \begin{figure}
\centering
\includegraphics[width = \columnwidth]{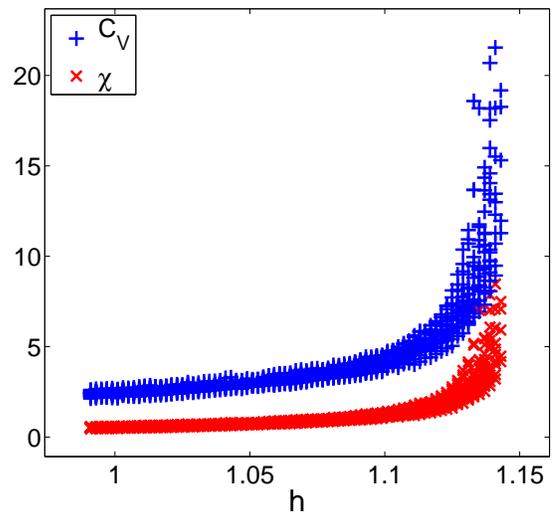}
\caption{(color online) The susceptibility and specific heat as a function of the distance from the spinodal for the dilute fully connected model. Twenty systems with a probability of a site being diluted of $0.1$ were simulated.
The existence of the spread for small $\Delta h$ is consistent with the theoretical analysis.}
\label{fig:Fig4}
\end{figure}

\begin{figure}
\centering
\includegraphics[width = \columnwidth]{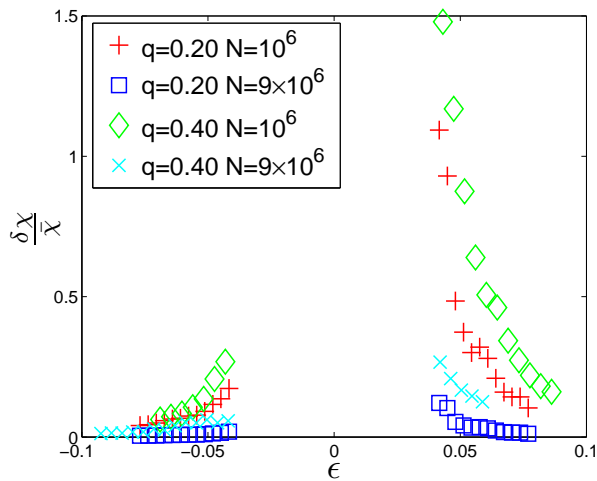}
\caption{(color online) Width of the susceptibility (variance normalized by the mean) distribution near the critical point (see Fig.~\ref{fig:Fig2})
for (a) $q=0.20$ and (b) $q=0.40$ with a cutoff of $\epsilon=0.04$.}
\label{fig:Fig5}
\end{figure}


In Fig.~\ref{fig:Fig4} we plot 20 realizations of the isothermal susceptibility and specific heat as a function of $h$ near the pseudospinodal for the fully connected model with $9\times 10^{4}$ spins. The empty sites are distributed with a probability of 0.10 as   for the critical point. Note that the curves for each simulation overlap until $\Delta h$ becomes too small at which point the pseudospinodal is smeared as evidenced by the fact that the curves show a distinct spread. This is consistent with our analysis. For both the critical point and the pseudospinodal the spread caused by the dilution in the fully connected model narrows for a given value of the distance from the pseudospinodal or the critical point as the system size is increased 
and hence $R$ increases as would be expected from our analysis.

In Fig.~\ref{fig:Fig5} we plot the width or spread of the susceptibility near the critical point (normalized by the susceptibility) for a $20\%$ and $40\%$ dilution in the fully connected model for two  system sizes. The width increases as $\epsilon$ decreases as expected. We did not display data for $\epsilon < 0.04$ because the nmf picture breaks down for $\epsilon \sim 0.05$. Similar results are obtained for the pseudospinodal~\cite{kks}.
Because the interaction range is a function of the system size, we expect the spread of the susceptibility associated with the dilution to decrease as the system size increases for the same value of $\epsilon$. This is consistent with the scaling analysis. A similar result is found near the spinodal~\cite{kks}.

In conclusion we have found that the effect of dilution in mf and nmf systems on critical points is more subtle than the effect in systems with short range interactions below the upper critical dimension, and it is different from the behavior of short range systems in $d=d_c$ dimensions. 
We also found that in mf systems the spinodal is not changed by dilution but in more physically realistic nmf systems the spread and consequent rounding of the pseudospinodal depends on the distance from the point of the apparent divergence. At the spinodal the HC is not simply related to the mf specific heat exponent~\cite{pelisetto}. Near $T_{c}$ for $h=0$ the Harris criterion is simply related to the specific heat for $T >T_{c}$. Near the critical point but for $T <T_{c}$ we found a result similar to what one finds near the spinodal. These results will be published in Ref.~\cite{kks}.

Finally, our results indicate that the effect of the spinodal on nucleation in a wide variety of systems is not eliminated by the presence of dilution and that spinodal nucleation  remains an important feature of systems with long range interactions.

Our results raise many questions and suggest several new research directions. What is the effect of dilution on nucleation and pseudospinodals in more complicated models associated with materials such as the long-range Ising antiferromagnet~\cite{klein2007}? What is the effect of other types of damage such as fixed spins or dislocations? What is the effect of various forms of damage and impurities in models in other fields such as econophysics and the spread of disease? The answers to these questions are of considerable importance in a variety of fields of study ranging from materials and biophysics to economics and social science.

\begin{acknowledgments}
This research was supported by the Department of Energy through grant DE-FG02-95ER14498. We acknowledge useful conversations with H.\ Gould, J.\ Silva and J.\ B.\ Rundle.
\end{acknowledgments}

\end{document}